\documentclass{appolb}
\usepackage{epsfig}
\usepackage{amsmath}
\def\be{\begin{eqnarray}}
\def\ee{\end{eqnarray}}
\def\bc{\begin{center}}
\def\ec{\end{center}}

\def\lsim{\stackrel{\scriptstyle <}{\phantom{}_{\sim}}}

\begin{document}
\pagestyle{plain}
\title{Open and hidden strangeness in hadronic systems%
\thanks{Presented at  Strangeness in Quark Matter,
September 18-24, 2011, Krak\'ow}%
}
\author{Boris Tom\'a\v{s}ik$^{a,b}$ and Evgeni E. Kolomeitsev$^{a}$
\address{$^a$Univerzita Mateja Bela, 97401 Bansk\'a Bystrica, Slovakia\\
$^b$Department of Physics, FNSPE, Czech Technical University in Prague\\11519 Prague 1,
Czech Republic}
}
\maketitle
\begin{abstract}
We investigate production of $\phi$ mesons and $\Xi$ baryons in nucleus-nucleus
collisions. Reactions on strange particles acting as a catalyser are proposed to
interpret the high observed $\phi$ yields in HADES experiments as well as the energy
dependence of the widths of $\phi$ rapidity spectra in collisions at the SPS energies. It
is argued that the enhancement of $\Xi^-$ yield observed by HADES is even higher than
originally reported if effects of the experimental centrality trigger are taken into
account. Cross sections for new hadronic processes that could produce $\Xi^-$ are
reviewed.
\end{abstract}


Interesting information about strangeness production and fireball dynamics in heavy ion collisions
is carried by $\phi$ mesons. The NA49 collaboration found puzzling collision energy dependence of
their rapidity spectra widths in Pb+Pb collisions at the CERN SPS \cite{PhiRMS}. The HADES
collaboration at the SIS machine of GSI Darmstadt, which investigates Ar+KCl collisions at
bombarding energies of 1.76~$A$GeV, has found strong enhancement of $\phi$
production~\cite{HADES-KKphi}. The same experiment has observed surprisingly high abundance of
doubly strange $\Xi^-$ baryon over $\Lambda$ hyperons~\cite{HADES-Xi}, which exceeds the
predictions of the statistical model~\cite{ABR06} as well as transport calculations~\cite{Chen04}.
In order to study strangeness enhancement it is essential to know how it is distributed among all
species and so a complete measurement of all species is needed; HADES did this
\cite{HADES-KKphi,HADES-Xi,HADES-Ko,HADES-Hyperons}. Here we report on recently proposed catalytic
$\phi$ meson production~\cite{KT09} which could help to explain the observations. We also
discuss comments about new effective mechanisms of $\Xi^-$ production.

Traditionally, for baryon-rich systems the main contribution to $\phi$ yield was thought to be
\emph{OZI suppressed} reactions like $\pi N \to \phi N$ \cite{nucl-th/9704002} and
\emph{stran\-ge\-ness coalescence} reactions $K\bar K\to \phi\rho$ and $K\Lambda\to \phi
N$~\cite{320155}. A new type of {\it catalytic reactions}
$$
\pi Y\to \phi Y\,, \qquad \bar{K} N\to \phi Y\,, \qquad Y=\Lambda\,,\, \Sigma\,.
$$
was proposed in~\cite{KT09}, in which only one rare strange reactant is needed and, furthermore,
the OZI suppression is lifted thanks to the presence of a strange particle acting as a
catalyser. Based on a simple hadronic effective theory we estimated and parameterized the
cross-sections of such processes to be typically of the order 1--2~mb~\cite{KT11}.

\begin{figure}
\centerline{
\parbox{5cm}{\includegraphics[width=4.5cm]{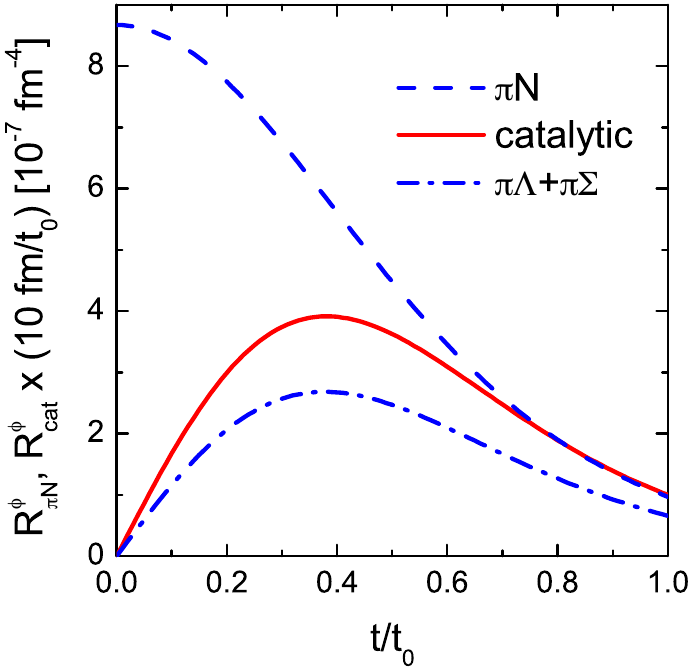}}\,\,
\parbox{5cm}{\includegraphics[width=4.5cm]{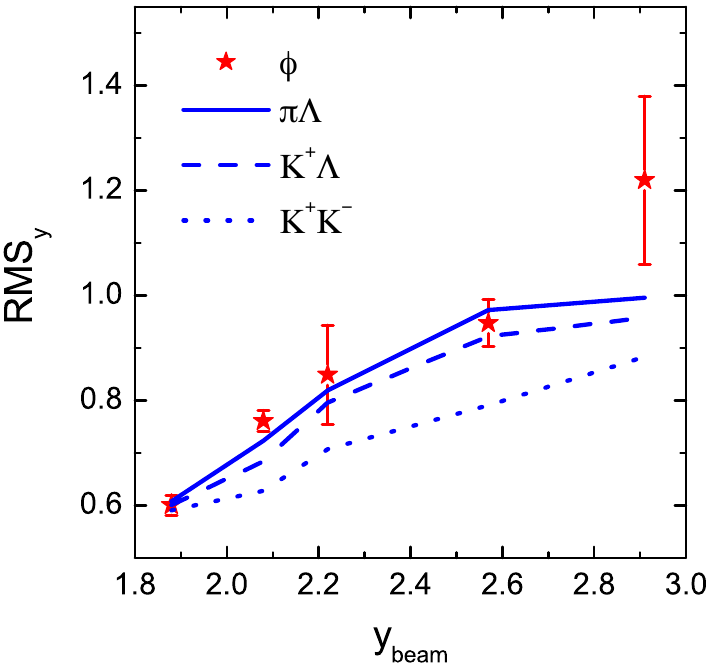}}
} \caption{Left panel: $\phi$ meson production rates. Solid
line is the sum of all catalytic reactions, dash-dotted line the reactions on hyperons.
Dashed line is the rate of the $\pi N\to \phi N$ reaction.
Right panel: root mean square of the rapidity distributions of $\phi$s produced in Pb+Pb
collisions versus the beam rapidity~\cite{PhiRMS}. Lines show the distribution widths from reactions
$\pi\Lambda\to \phi Y$, $K^+\Lambda\to \phi N$ and $K^+K^-\to \phi$.}
\label{fig:FRates}
\end{figure}

In order to see their efficiency we set up a kinetic calculation
assuming a simple parameterization for the time dependence of the net baryon density and
temperature. Strangeness was taken into account perturbatively. In Fig.~\ref{fig:FRates}
(left panel)
we compare the production rates of catalytic reactions with those of $\pi N \to \phi N$
processes. For reaction $a+b\to c+d$ its rate is determined as
$R^{cd}_{ab} = \rho_a \rho_b \langle \sigma^{cd}_{ab} v_{ab}\rangle / (1 + \delta_{ab})$
where $\rho$'s are the densities of reactants and $\langle \sigma^{cd}_{ab} v_{ab}\rangle$
is the cross-section multiplied by relative velocity and averaged over the distribution
of velocities. The density evolution in that calculation corresponds
roughly to collisions at beam energy of 6~$A$GeV with total fireball lifetime of 10~fm/$c$
(for more details see \cite{KT09,KT11}). We see that the contribution of catalytic reactions starts
gradually as the catalysing strange hadrons are being produced and then becomes comparable
to that of the usually assumed processes.

Considering the rapidity distribution of $\phi$ mesons one basically assumes that it is
given by the product of rapidity distributions of those species which produce $\phi$s,
provided their rapidity distributions would not change in the course of the
collision. The observed widths of $\phi$ rapidity distributions do not follow, however,
from the product of kaon rapidity distributions, although kaon coalescence was supposedly
the dominant process \cite{320155}. Instead, the reasonable description of the energy dependence of
$\phi$ rapidity widths is obtained if one combines mesons with $\Lambda$ hyperons---like
it would follow from the catalytic reactions. Deviation from such an interpretation observed
only at the collision energy of 158~$A$GeV might indicate that here another mechanism of
$\phi$ production sets in. This is an interesting observation in view of the currently
running energy-scan programme and the search for the onset of deconfinement.

As measured by the HADES collaboration \cite{HADES-Xi} the yield of the doubly strange
$\Xi$ hyperons is dramatically enhanced in comparison to statistical model predictions.
We have studied ratios of strange species with the constraint that the total density of
all $S<0$ hadrons is set by the measured number of $K^+$ mesons. In our calculation
strange quarks are distributed among hadrons according to the statistical model with
temperature as a free parameter. In this treatment the density
of species $i$ is given as $\rho_i = \lambda_S^{s_i} n_i(T,\mu_B)$, where $n_i$ is the
thermal density at the temperature $T$ and chemical potential $\mu_B$, and the
strangeness fugacity $\lambda_S$ is proportional to the density of strange quarks; $s_i$
is strangeness of species $i$. Note that the cascade number depend quadratically on the density
of strange quarks. This is important because the reported tenfold enhancement of $\Xi$'s
\cite{HADES-Xi} does not yet take into account the LVL1 trigger used by HADES to get rid
off peripheral collisions. In the collisions selected by this trigger the fireball volume
is effectively larger than the volume averaged over minimum bias events. Consequently,
the density of strange quarks is lower and the expected concentration of cascades is also
lower. In summary, the observed enhancement of $\Xi$ production is even stronger than
reported originally, if the centrality selection is properly accounted for. We checked
that this observation holds qualitatively even if the masses of the involved hadrons
change in medium to the maximum possible extent. Therefore, one cannot exclude some
mechanism of an out-of-equilibrium production of $\Xi$ baryons in the nuclear collision.

\begin{figure}
\centerline{
\parbox{5cm}{\includegraphics[width=4.5cm]{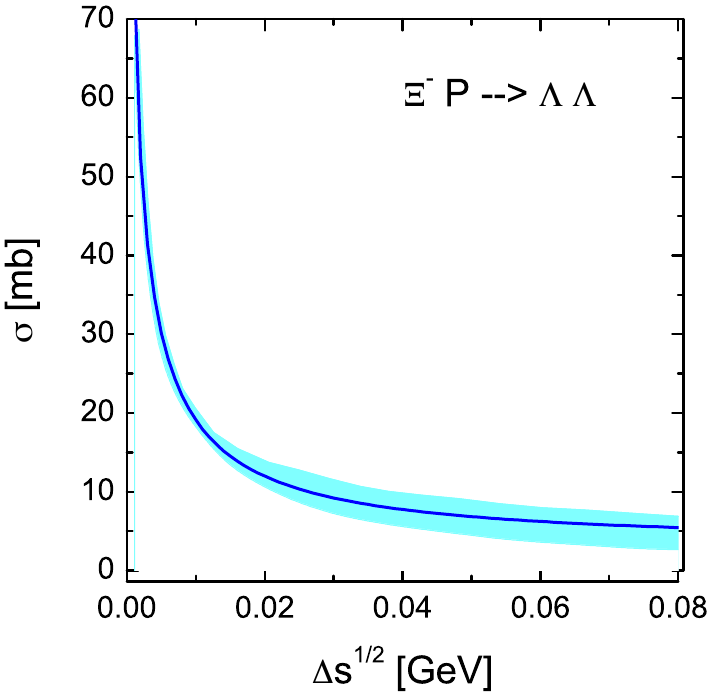}}\quad
\parbox{5cm}{\includegraphics[width=4.5cm]{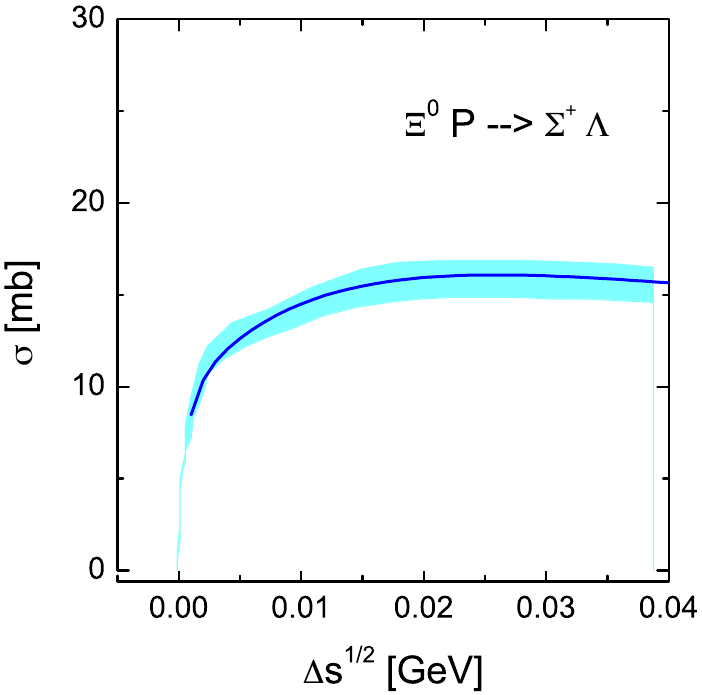}}}
\caption{
The cross sections  for the reactions
$\Xi^- p\to \Lambda\Lambda $ and $\Xi^0 p\to \Sigma^+ \Lambda$.
Shadowed bands are calculations from \cite{Polinder:2007mp}.
Solid lines are the parameterizations (\ref{Mfit-LL}) and (\ref{Mfit-SL}).}
\label{Fig:S-Sinv}
\end{figure}

Let us now review the processes, in which $\Xi$ can be produced. Firstly, there are {\it
strangeness creation} reactions:
$\bar{K} N\to K \Xi - \mathrm{380~{MeV}}$,
$\pi\Sigma \to K \Xi  - \mathrm{480~{ MeV}}$,
$\pi\Lambda \to K \Xi - \mathrm{560~{ MeV}}$.
They are endotermic, and their rates are strongly suppressed at temperatures $T\lsim 100$~MeV 
reachable at SIS energies.
Secondly, there are exothermic {\it strangeness recombination} reactions. They
can be anti-kaon-induced:
$\overline{K}\Lambda(\Sigma) \to  \Xi\pi + \mathrm{154(232)~{ MeV}}$
with the cross-sections calculated in \cite{LiKo02} and routinely included in transport 
codes~\cite{Chen04}.
The other group or reactions, which have not been included so far in transport codes,
are {\it double-hyperon processes}
$\Lambda\, \Lambda \to \Xi N- \mathrm{26~{ MeV}}$,
$\Lambda\, \Sigma \to \Xi N + \mathrm{52~{ MeV}}$,
$\Sigma\, \Sigma \to \Xi N+ \mathrm{130~{ MeV}}$.
At SIS energies the yields of hyperons are
an order of magnitude higher than those of antikaons, so we expect higher contribution from the
double-hyperon processes.
The cross-section of such $\Xi^-$ production processes can be obtained from the cross sections
of $\Xi^- p\to \Lambda\Lambda $ and $\Xi^0 p\to \Sigma^+ \Lambda$ reactions
calculated in \cite{Polinder:2007mp} within the chiral effective field theory.
We parametrize the matrix elements as
\begin{equation}
|M_{\Lambda\Lambda}|^2 = \big[ 3 + 12/( 1+80\, x_1)\big]~{\rm mb} \,,
\label{Mfit-LL}
\end{equation}
for $\Xi^- p \to \Lambda \Lambda$, where $x_1=(\sqrt{s}-m_{\Xi^-}-m_p)/{\rm 1~GeV}$, and
for $\Xi^0 p\to \Sigma^+ \Lambda$
\begin{equation}
|M_{\Sigma\Lambda}|^2 =
\big[ 5.5 + 36/(1 + 20\, x_2)^{1.2} + 32/(1 + 100\, x_2)^5\big]~{\rm  mb}\,,
\label{Mfit-SL}
\end{equation}
where $x_2=(\sqrt{s}-m_{\Sigma^+}-m_\Lambda)/{\rm 1~GeV}$.
Quality of these parameterizations is demonstrated in Fig.~\ref{Fig:S-Sinv}.
Cross-sections for the inverse reactions are calculated just by using proper
pre-factors
\begin{equation}
\sigma_{\Lambda\Lambda\to \Xi^- p}(s) =
(p_{p\Xi}/p_{\Lambda\Lambda})\, |M_{\Lambda\Lambda}|^2,\quad
\sigma_{\Sigma^+ \Lambda\to \Xi^0 p}(s) =
(p_{p\Xi}/p_{\Sigma\Lambda}) \, |M_{\Sigma\Lambda}|^2\, ,
\end{equation}
where $p_{ab}(\sqrt{s})$ is the CMS momentum of the pair of hadrons $a$ and $b$ with
total energy $\sqrt{s}$. For the other isospin channels we have
$\sigma_{\Lambda\Lambda\to \Xi^0 n}=\sigma_{\Lambda\Lambda\to \Xi^- p},$
\begin{align*}
\sigma_{\Sigma^0 \Lambda\to \Xi^- p}&={\textstyle\frac12}
\sigma_{\Sigma^+ \Lambda\to \Xi^0 p}, &
\,\,
\sigma_{\Sigma^0 \Lambda\to \Xi^0 n}& ={\textstyle\frac12}
\sigma_{\Sigma^+ \Lambda\to \Xi^0 p},
\\
\sigma_{\Sigma^- \Lambda\to \Xi^- n}& =
\sigma_{\Sigma^+ \Lambda\to \Xi^0 p}.
\end{align*}

\begin{figure}
\centerline{
\includegraphics[width=4.5cm]{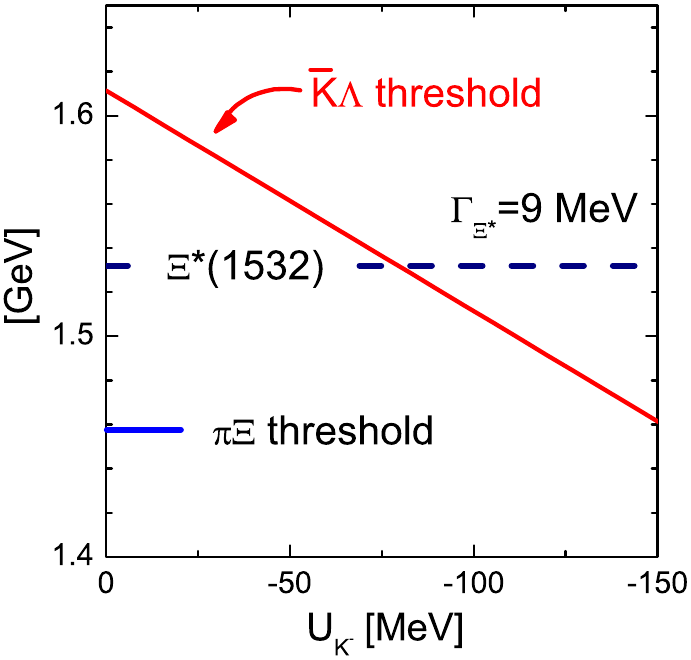}
\quad
\includegraphics[width=4.5cm]{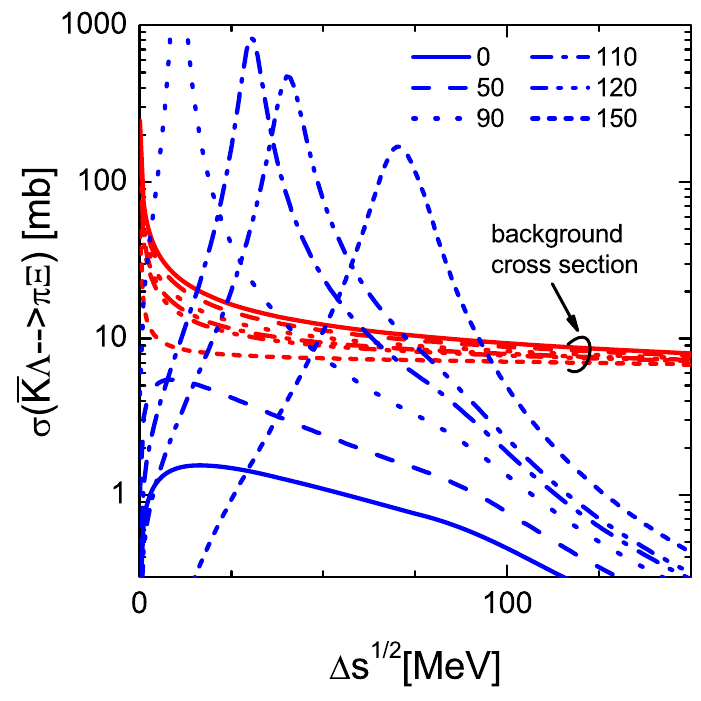}
} \caption{
Left: Threshold energy for $\overline{K}\Lambda$-induced reaction as a
function of attractive kaon potential. The dashed line shows the position of the $\Xi^*$
pole. Right: Resonant cross-section (\ref{e:genxs}) versus
$\Delta s^{\mbox{\tiny 1/2}}=s^{\mbox{\tiny 1/2}}-m_\Lambda-m_K-U_{K^-}$; the indices show the values of the
attractive antikaon potential $(-U)$ in MeV. For comparison background cross-section
(\ref{KL-cross}) is also shown in the same figure.
} \label{fig:Rescross}
\end{figure}

Let us now turn back to reactions $\bar K Y \to \Xi \pi$. Their cross-sections
were calculated in a coupled-channels approach in \cite{LiKo02}.
The isospin-averaged cross-sections are parameterized in \cite{Chen04} as
\be
&&\sigma_{\overline{K}\Lambda\to \pi\Xi} =
\frac{\phantom{4}\,p_{\pi\Xi}}{4\,p_{\overline{K}\Lambda}}\,
34.7\, \frac{s_{\overline{K}\Lambda}}{s}\,{\rm mb}\,,
\label{KL-cross}\\
&&\sigma_{\overline{K}\Sigma\to
\pi\Xi} = \frac{\phantom{12}\,p_{\pi\Xi}}{12\,p_{\overline{K}\Sigma}}\,
 318\,\big[1-\frac{s_{\overline{K}\Sigma}}{s}\big]^{0.6}\,
\big[\frac{s_{\overline{K}\Sigma}}{s}\big]^{1.7}
\,{\rm mb}\,.
\label{KS-cross}
\ee
where the symbol $p_{ab}$ was introduced above, and the reaction threshold is
$s_{\mbox{\tiny $\overline{K} Y$}} = ( m_K + m_{Y})^2$. The change of the cross section
in medium due to a lowering kaon mass is usually taken into account \cite{Chen04,LLB97} by
shifting the reaction threshold $\sigma^*(\sqrt{s})=\sigma(\sqrt{s}+s^{\mbox{\tiny
1/2}}_{\mbox{\tiny $\overline{K} Y$}} -s^{*,\mbox{\tiny 1/2}}_{\mbox{\tiny $\overline{K}
Y$}})$, where $s^{*}_{\mbox{\tiny $\overline{K} Y$}}$ is the reaction threshold with the
in-medium particle masses. Doing so one misses a potentially important effect. If kaon
mass is lowered enough, the resonant process $\overline{K}Y \to \Xi^*(1530) \to \Xi \pi$
not included in~\cite{LiKo02} may become relevant. In Fig.~\ref{fig:Rescross} (left
panel)
we show that the reaction threshold crosses the resonance pole at the $\overline{K}$ potential
$U_{K^-} \simeq - 80\, \mbox{MeV}$ The cross-section of the resonant reaction can be simply added
to the parameterizations (\ref{KL-cross}) and (\ref{KS-cross}). Overall structure of the
cross-section for the resonance process is given by
\begin{equation}
\label{e:genxs}
\sigma_{\rm res}(\sqrt{s})=({2\, \pi}/{p_{\mathrm{in}}^2})\,
{\Gamma_{\rm in}\, \Gamma_{\mathrm{out}}}[{(\sqrt{s}-m_{\rm res})^2+
\Gamma^2_{\rm tot}/4}]^{-1},
\end{equation}
where $\Gamma_{\rm tot}$ is the total width of the resonance and
$\Gamma_{\mathrm{in}}$ ($\Gamma_{\mathrm{out}}$) is the partial width
in the in(out)going channel. The partial widths for the decay of a
${\frac{3}{2}}^+$ resonance with rest energy $\sqrt{s}$ into a ${\frac{1}{2}}^+$ baryon
with the mass $m_B$ and $0^-$ meson with the mass $m_M$ can be written
(see e.g.\ page 310 in \cite{Gasior})
\begin{equation}
\Gamma_{MB}(\sqrt{s})=\frac{V^2\, p^3}{12\, \pi\, \sqrt{s}}\,
\big[(p_{MB}^2+m_B^2)^{1/2}+m_B\big]\theta(\sqrt{s}-m_B-m_M),
\label{Ggen}
\end{equation}
where $V$ is the coupling constant of the order $O(m_\pi^{-2})$. From the measured value of the
decay width $\Gamma(\Xi^*\to \Xi\pi)=9$~MeV, we find $V_{\Xi^* \Xi\pi}=1.315/m_\pi^2$. The SU(3)
symmetry implies that 
$V_{\Xi^* \Lambda \bar{K}}\simeq V_{\Xi^* \Sigma \bar{K}}\simeq V_{\Xi^*\Xi\pi}$. 
In order to get isospin-averaged cross-section for the $\Sigma$ induced reactions,
Eq. (\ref{e:genxs}) is to be multiplied by factor 1/3. The total width of $\Xi^*$ is
given by the sum of all partial widths
$
\Gamma_{\rm tot}(\sqrt{s})=\Gamma_{\pi\Xi}(\sqrt{s})+
\Gamma_{\bar{K}\Lambda}(\sqrt{s}) + \Gamma_{\bar{K}\Sigma}(\sqrt{s})
$.

The resonant cross-section for various values of the kaon potential is compared with the
background cross-section (\ref{KL-cross}) in Fig.~\ref{fig:Rescross} (right panel). We conclude
that for selected energies above the threshold the contribution of the resonant reaction may be
significant.

The observed enhanced yields of $\phi$s and $\Xi$s  most likely indicate a non-equilibrium
production mechanism. This may occur in an expanding fireball if the rates for annihilating given
species cannot follow the fast drop of temperature.

This work has been supported in part by
MSM~6840770039, and  LC~07048 (Czech Republic). BT acknowledges support of the
Slovak-Polish collaboration grant No SK-PL-0021-09.


\end{document}